\documentclass{ws-p8-50x6-00}

\def\gs{\mathrel{\raise0.35ex\hbox{$\scriptstyle >$}\kern-0.6em
\lower0.40ex\hbox{{$\scriptstyle \sim$}}}}
\def\ls{\mathrel{\raise0.35ex\hbox{$\scriptstyle <$}\kern-0.6em
\lower0.40ex\hbox{{$\scriptstyle \sim$}}}}

\begin{document}

\title{The Nature of Faint Submillimeter Galaxies}

\author{Ian Smail}

\address{Department of Physics, University of Durham, South Road,
        Durham DH1 3LE\\E-mail: ian.smail@durham.ac.uk}

\author{Rob Ivison}

\address{Department of Physics \& Astronomy, University College London, 
	Gower Street, London WC1E 6BT
	E-mail: rji@star.ucl.ac.uk}  

\author{Andrew Blain}

\address{Institute of Astronomy, Madingley Road, Cambridge CB3 0HA\\
	E-mail: awb@ast.cam.ac.uk}  

\author{Jean-Paul Kneib}

\address{Observatoire de Toulouse, 14 avenue E.\ Belin,
        31400 Toulouse, France\\
	E-mail: jean-paul.kneib@ast.obs-mip.fr}  

\maketitle

\abstracts{
We summarise the main results on the faint submillimeter (submm) galaxy
population which have come from the SCUBA Cluster Lens Survey.  We
detail our current understanding of the characteristics of these
submm-selected galaxies across wavebands from X-rays to radio.
After presenting the main observational properties of this population
we conclude by discussing the nature of these distant, ultraluminous
infrared galaxies and their relationship to other high-redshift populations.
}

\section{Introduction}

The results of the highly successful far-infrared (FIR) survey
undertaken by  {\it IRAS}  led to a wide-spread realisation of the
ubiquity and importance of highly-obscured star-forming and active
galaxies in the local universe.  More recent work in the FIR and submm
wavebands has produced a similar revolution in our understanding of
obscured galaxies in the distant universe.  These observations employ
the {\it COBE} and {\it ISO} satellites and the  Sub-millimeter Common
User Bolometer Array (SCUBA\cite{scuba}) on the 15-m JCMT\footnote{The
JCMT is operated by the Joint Astronomy Centre on behalf of the United
Kingdom Particle Physics and Astronomy Research Council (PPARC), the
Netherlands Organisation for Scientific Research, and the National
Research Council of Canada.} and have shown that obscured galaxies
contribute a substantial fraction of the total emitted radiation at
high redshifts.

The rough equivalence of the energy density in the optical
background and that detected in the FIR/submm  by {\it
COBE}  suggests that, averaged over all epochs, approximately
half of the total radiation in the universe came from obscured sources
(either stars or AGN). Clearly including these class of galaxies in
models of galaxy evolution is critical to obtain a complete
understanding of the formation and evolution of galaxies.

As we will show in the next section, the bulk of the emission in the
FIR/submm comes from a relatively small population of extremely
luminous, dusty galaxies.   These galaxies lie at high redshifts, $z\gs
1$--4, are both massive and gas-rich and they may dominate the total
star formation in the universe at these early epochs.  The analogs of
this population in the local universe are the Ultra-Luminous Infrared
Galaxies (ULIRGs) uncovered by {\it IRAS}.  As a benchmark for the
following discussion we note that a ULIRG similar to Arp\,220 with a
far-infrared luminosity of $L_{FIR} \sim 3 \times 10^{12} L_\odot$ and
a star-formation rate (SFR) of $\sim 300$\,M$_\odot$ yr$^{-1}$ would
have a 850-$\mu$m flux density of $\gs 3$\,mJy out to $z\sim 10$ in a
spatially flat Universe.\footnote{We assume $q_o=0.5$ and $h_{\rm
100}=0.5$ unless otherwise stated.}

\section{Submm number counts and the FIR background}

The advent of sensitive submm imaging with SCUBA has allowed a number
of groups to undertake `blank'-field surveys for faint submm galaxies.
Results on the number density of sources in blank fields as a function
of 850-$\mu$m flux density have been published by three
groups.\cite{hdf,sae99,ajb98}   Unfortunately, due to the modest
resolution of the SCUBA maps, 15$''$ FWHM, these surveys are confusion
limited at $\sim 2$\,mJy and resolve $\sim 50$\% of the {\it COBE} background.

Our collaboration has taken a complimentary approach to these `blank'
field surveys by using massive gravitational cluster lenses to increase
the sensitivity and resolution of SCUBA.   This survey covers seven
lensing clusters at $z=0.19$--0.41\cite{sib,sibk,bkis} (new results
from two similar lensing surveys with SCUBA were also reported at this
meeting\cite{pvdw,dssc}).  Our analysis  uses  well-constrained lens
models to  correct the observed source fluxes for lens
amplification.\cite{bkis}  For the median source amplification, $\sim
2.5\times$, this survey covers an area of the source plane equivalent
to  15\,arcmin$^2$ at a $3\sigma$ flux limit of $\sim 2$\,mJy.  The
amplification also results in a factor of two finer beam size at this
depth so that these counts have a fainter confusion limit than the
blank field observations.  The surface density of sources at the limit
of our survey (0.5\,mJy) is $\sim 8$\,arcmin$^{-2}$ (the equivalent
density of `normal' field galaxies is reached at $I\sim 23.5$) and we
resolve $\sim 100$\% of the {\it COBE} background.\cite{bkis} We
conclude that the majority of the extragalactic submm background arises
in a relatively sparse population of extremely luminous galaxies, $L\gs
10^{12} L_\odot$.

\section{Optical counterparts to submm galaxies}

\begin{figure}[th]
\vspace*{0.8in}
\centerline{\Huge fig1.gif}
\vspace*{0.8in}
~\bigskip
\caption{ 
$15''\times 15''$ images of the 17 sub-mm sources in our full
sample.  These are ordered from the upper-left on the basis of the
reliability of their proposed optical counterparts -- with the first
two rows containing those sources which were probably correctly
identified in the optical imaging.$^{21,25}$ Possible counterparts are
identified with `?', those which are unlikely to be the correct
counterpart are marked with an `X'.  The bottom row shows those submm
source where subsequent near-infrared and radio analysis has brought
into question the identification of the proposed optical counterparts.
}
\end{figure}

\vspace*{-0.5cm}
The first attempts at identifying counterparts to the submm sources in
our survey concentrated on exploiting the high-quality, archival {\it
Hubble Space Telescope} {\it WFPC2} imaging which existed for the
majority of these fields.\cite{sibk}  These frames typically reach
limits of $I\sim 27$ on the background source-plane and we felt
confident of identifying a large fraction of the submm galaxies
(reflecting the optical bias of the author).  Figure~1 shows the
identifications from Smail et al.\ (1998)\cite{sibk} but ordered in
terms of our current understanding of the reliability of the proposed
counterparts (based on the near-infrared and radio follow-up discussed
in the next section).

If we remove the two central cluster galaxies\cite{e99} from our sample
(SMM\,J21536+1741 and SMM\,J14010+0252)  --  we conclude that only
about half of the original identifications which were proposed are
likely to be correct.  Indeed, of all the  submm galaxies detected in
all the SCUBA surveys, only three have confirmed counterparts:
SMM\,J02399$-$0136,\cite{i98} SMM\,J14011+0252\cite{i00} and
SMM\,J02399$-$0134 (Fig.~1), where the confirmation comes from
identification of CO emission at the same redshift as the proposed
counterpart.\cite{f98,f99,k01}  For a handful of other sources, their
continuum detections with millimeter interferometers and the improved
astrometry which these provide, have been used to identify likely
counterparts.\cite{g00,c00} Equally, identifications of counterparts
have been claimed when a galaxy with sufficiently unusual
characteristics is found within the submm error box (e.g.\ an
ERO\cite{s99}), but so far none of these have been confirmed in CO.

The relatively low identification rate, even with the high-quality
optical imaging we have available, results partly from the fact that
source confusion with the large SCUBA beam is compounded by clustering,
potentially leading to deviations of the order of the
beam-size.\cite{h00} Indeed the clearest route to identifying
counterparts is usually to overlay the raw SCUBA map onto the optical
image and to use the information present in the shape of the submm
emission to search for possible merging of sources.  We believe that
source confusion may have contributed to the ambiguities in the
identification of counterparts for SMM\,J21536+1742, SMM\,J22471$-$0206
and SMM\,J02400$-$0134.  However, the main difficulty with identifying
optical counterparts arises because the majority  of the submm population
are {\it extremely} faint at optical wavelengths: at least half and
perhaps upto 75\% of galaxies in Fig.~1 have optically faint
counterparts, $I\gg 25$--27 (i.e.\ Classes~0 and 1\cite{i01}).  Removing the
known AGN from the sample, the proportion of starburst-powered submm
galaxies which have optically faint counterparts rises to 90\%.  We now
briefly discuss some of the observations which have led us to this
conclusion.

\section{Near-infrared and radio counterparts to submm galaxies}

The first indication that things might not be as clear-cut as the
proposed optical identifications suggested came from $K$-band imaging
of our SCUBA fields using UKIRT\footnote{UKIRT is operated by the Joint
Astronomy Centre on behalf of PPARC.}  These images reach $K\sim 21$ in
the source plane and uncovered  new candidates (H5 and N4 in Fig.~2)
within the submm error-boxes of two of the submm sources.\cite{s99}
These galaxies are undetected in very deep {\it HST} and Keck $R$- and
$I$-band imaging ($I\gs 27$ corrected for lens amplification), but are
relatively bright in the $K$-band.  Their colours, $(I-K)\gs 6.0$ and
$\gs 6.8$ for H5 and N4 respectively, place them firmly in the rare
class of Extremely Red Objects (EROs).\cite{s99} The surface density of
ERO-submm galaxies in our survey means that they account for around
half of all ERO's -- providing an important link between these two
populations. Moreover, the extreme submm to optical ratios of these
galaxies, $L_{FIR}/L_{B} \gs 300$ (assuming they lie at $z\sim 2.5$--3,
as suggested by fits to their SEDs), combined with their high inferred
star formation rates underlines the difficulties faced when attempting
to use UV-selected samples to audit  star formation at $z\gg 1$.
\vspace*{-0.2cm}

\begin{figure}[th]
\centerline{\psfig{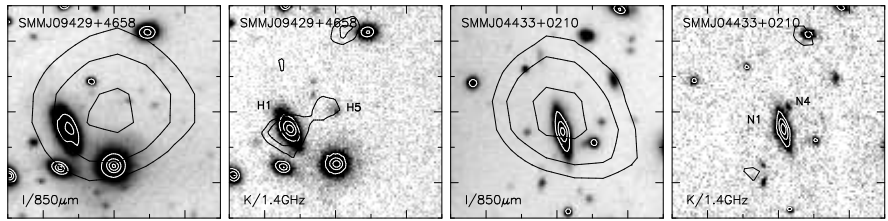}} 
\caption{
ERO counterparts to two submm sources in our survey.$^{18}$ The two
panels for each source show deep $I$- and $K$-band images with the
850\,$\mu$m and 1.4\,GHz maps overlayed.  The faintest sources visible
in the $I$-band exposure have observed magnitudes of $I\sim 25.5$--26.0
and $K\sim 20.5$.  The original counterparts proposed for the submm
sources are marked on the $K$-band images, as well as the new ERO
candidates, H5 for SMM\,J09429+4658 and N4 for SMM\,J04433+0210.  Each
panel is 30$''$ square and is centred on the 850-$\mu$m position.
}
\end{figure}

\vspace*{-0.5cm}
The next phase of the identification process used {\it very} deep
1.4\,GHz maps of our fields from the VLA.\cite{s00}  These maps allow
us to identify radio counterparts or place stringent limits ($\ls
20\mu$Jy in the source plane) on the radio flux of the submm galaxies.
Moreover, the recent work on submm-to-radio spectral indices for
distant star forming galaxies\cite{cy99,cy00,awb} has provided us
with a useful tool for estimating  the redshift distribution of a {\it
complete} submm-selected sample.\cite{s00} We can compare the redshift
limits derived in this manner with the spectroscopic redshifts for
individual candidate optical counterparts\cite{b99} and hence determine
the reliability of these proposed counterparts as shown in Fig.~1.  On
the basis of this analysis we conclude that the submm population
brighter than $\sim 1$\,mJy has a median redshift of {\it at least}
$<\!\! z\!\! >\sim 2$, and using the available spectral index models,
probably $<\!\! z\!\! >\sim 2.5$--3.\cite{s00} The high median redshift
means that the submm population, if predominately powered by starbursts
(rather than AGN), contributes a substantial fraction of the total star
formation density at high redshifts.
   
Our on-going program of observations of this sample of submm galaxies,
using near-infrared imaging and spectroscopy and millimeter
interferometry is reported elsewhere in this volume.\cite{dtf}

\section{The submm population at other wavelengths}

To understand the role of the submm galaxies in galaxy formation and
evolution we first have to determine the relative contributions to
their extreme luminosities from AGN and starburst-powered emission.
Hard X-ray observations can uncover the presence of a dominant AGN
power source in even highly obscured sources, and hence provide a
simple estimate of the AGN fraction in the submm
population.\cite{omar,kfg} Unfortunately, even for high-redshift
sources, {\it Chandra} provides relatively poor hard X-ray response,
and hence to-date all the X-ray studies of submm
galaxies\cite{acf,a370} have achieved is to confirm the presence of
modestly-obscured AGN in those galaxies in which optical spectroscopy
had already identified AGN emission.\cite{i98,b99} These galaxies
account for 10--20\% of the submm population, in-line with estimates
from the models of the X-ray background.\cite{omar,kfg}  Much more
interesting limits on the fraction of highly-obscured AGN in this
population will come from {\it Newton} observations.  Equally useful
constraints on the fraction of obscured AGN and the breakdown of AGN-
versus starburst-powered emission in individual galaxies will come from
high-resolution, mid-infrared observations (using OSCIR or Michelle on
Gemini) which are sensitive to emission from hot-dust and PAHs.
However, we conclude that, on present evidence, it seems safe to assume
that the majority of the emission from the submm population is powered
by star formation.\cite{bsik}

\section{The relationship between FIR- and UV-selected galaxies}

Finally, we briefly discuss  the relationship of submm galaxies to
other classes of high-redshift sources.  Recently it has been claimed
that UV- and optically-selected galaxy samples at $z\sim 0$--3 can
completely account for the extragalactic background, not only in the
optical/UV, but also in the far-infrared and submm.\cite{kacc} This
calculation relies upon the correction of the UV luminosities of these
samples for the effects of dust extinction, using a correlation between
reddening (measured through the spectral slope in the UV) and UV
luminosity from observations of normal galaxies at low redshifts.
Unfortunately, as was graphically illustrated at this
conference,\cite{ds} this correlation doesn't hold (even locally) for
galaxies with bolometric luminosities typical of the population we have
found at higher redshifts which dominate the submm counts and produce
the bulk of the {\it COBE} background.  This clearly calls into
question the reliability of this type of calculation.

The argument that UV-selected samples can provide a complete picture is
supported using optical/UV and submm observations of an example of a
`representative' submm galaxy, SMM\,J14011+0252 (Fig.~3).\cite{i00}  However,
this galaxy is far from typical of the majority of the submm
population  -- which would be better described using the ERO, HR\,10,
as an archetype.\cite{d99} Moreover, even within the SMM\,J14011+0252
system it appears from interferometry observations that the dominant
bolometric component resides, not in the UV-bright companion J2, but in
the much redder galaxy J1.  Using a UV-weighted spectral slope (which
has a substantial contribution from J2) in an attempt to predict the
unobscured star formation rate (which mostly occurs in J1) in this
system seems questionable -- why should the UV properties of J2 have
any bearing on the degree of obscuration and star formation in J1?
\vspace*{-0.2cm}

\begin{figure}[th]
\centerline{\psfig{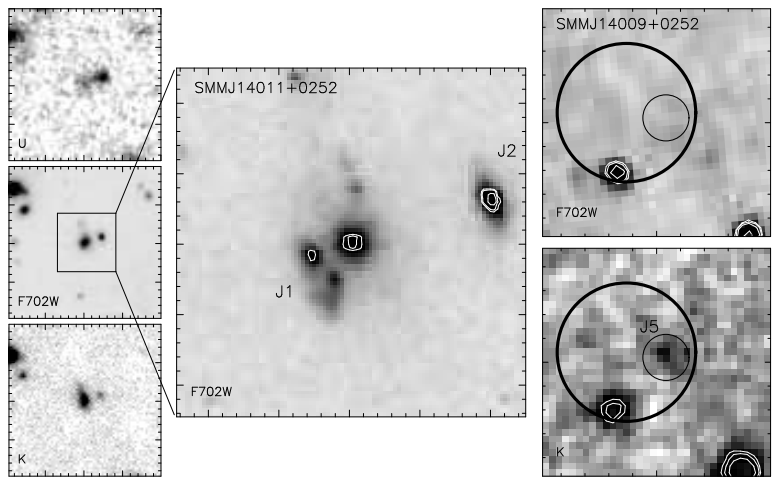}}
\caption{
These panels show two of the submm galaxies detected through the
lensing cluster A\,1835.$^{12}$ The left-hand panels illustrate the
morphology of the two components, J1/J2, of the $z=2.55$ starburst
galaxy SMM\,J14011+0252, from the $U$- to $K$-bands.  The central panel
gives a zoomed view of the {\it WFPC2} $R$-band image of this galaxy,
demonstrating the complex, merger-like morphology of the
bolometrically-dominant component J1.  The right-hand panels show $R$-
and $K$-band views of a more typical submm galaxy, SMM\,J14009+0252,
for which a very faint near-infrared companion, J5, has been
identified.$^{12}$ This galaxy is clearly undetected in the deep {\it
HST} $R$-band exposure.  The major tickmarks are every 5$''$, except
for the central panel were they are every 1$''$.
}

\end{figure}

\vspace*{-0.5cm}
Instead, if we adopt HR\,10 as a template for a typical submm galaxy
then we find that:  {\it as HR10 has a far higher ratio of far-IR to
far-UV luminosity than the other galaxies considered it would not be
included in UV-selected surveys despite its large star formation
rate.}\cite{kacc}  We conclude therefore that UV light is an
inappropriate tracer of star formation in the highly-obscured galaxies
that contain a large fraction of the star formation in the high
redshift universe.  Further study of this remarkable population is essential
for a complete understanding of the formation of massive galaxies.\cite{bsik,i01}

\vspace*{-0.3cm}
\section*{Acknowledgments}
We thank our collaborator on the {\it HST} lensing project, Harald
Ebeling, for allowing us to present the {\it WFPC2} image of A\,1835.
IRS acknowledges travel support to attend this conference.  The
research presented here was supported by the Royal Society [IRS], PPARC
[RJI],  the Raymond and Beverly Sackler Foundations [AWB] and the CNRS
[JPK].

\end{document}